\documentclass[prd,twocolumn,superscriptaddress,floatfix,nofootinbib]{revtex4-1}
\usepackage[utf8]{inputenc}
\usepackage{graphicx}
\usepackage{graphics,bm}
\usepackage{amssymb}
\usepackage{amsmath}
\usepackage{mathtools}
\usepackage{color}
\usepackage{multirow}
\usepackage{subcaption}

\newcommand{\bra}[1]{\langle #1 |}
\newcommand{\ket}[1]{\left| #1 \right\rangle}

\def\6{\langle}
\def\9{\rangle}

\usepackage{amsmath}
\usepackage{amsfonts}
\usepackage{amssymb}

\newcommand\tb{\mathtt{b}}

\newcommand\tC{\mathtt{C}}
\newcommand\tS{\mathtt{S}}
\newcommand\tN{\mathtt{N}}
\newcommand\tM{\mathtt{M}}
\newcommand\tL{\mathtt{L}}

\newcommand\ttt{\mathtt{t}}
\newcommand\tx{\mathtt{x}}
\newcommand\tE{\mathtt{E}}
\newcommand\tp{\mathtt{p}}
\newcommand\tw{\mathtt{w}}

\newcommand\tu{\mathtt{u}}

\newcommand\wtPsi{\widetilde{\mathtt{\Psi}}}

\newcommand{\be}{\begin{equation}}
\newcommand{\ee}{\end{equation}}
\newcommand{\ba}{\begin{eqnarray}}
\newcommand{\ea}{\end{eqnarray}}

\usepackage{amsmath}

\begin{document}
\title{Entangled particle-states localized on curved space-time}

\author{Vasileios I. Kiosses}
\email{kiosses.vas@gmail.com}
\affiliation{QSTAR, INO-CNR, Largo Enrico Fermi 2, I-50125 Firenze, Italy}

\begin{abstract}
	In this work, based on a recently introduced localization scheme for scalar fields, we argue that the geometry of the space-time, where the particle states of a scalar field are localized, is intimately related to the quantum entanglement of these states. More specifically, we show that on curved space-time can only be localized entangled states, while separable states are located on flat space-time.
	Our result goes in parallel with recent theoretical developments in the context of AdS/CFT correspondence which uncovered connections between gravity and quantum entanglement.
\end{abstract}

\maketitle

\noindent

\section{Introduction}

If quantum mechanics, as it is broadly believed, is the underlying framework for our physics, then eventually gravity should fit into this quantum mechanical context.
Quantum field theory, founded on quantum mechanics, already describes with great success the other fundamental interactions through the standard model of particle physics.

Einstein's legacy in understanding gravity is the realization that space-time and gravitational field are the same physical entity. However, the idea of obtaining a quantum version of gravity by applying the standard quantization rules directly to the space-time metric runs into insurmountable problems.

A big theoretical step forward in understanding quantum gravity has come from AdS/CFT correspondence \cite{Maldacena}. It postulates that ordinary quantum field theories are exactly equivalent to certain quantum theories of gravity with specific space-time asymptotic behavior. However, despite the large amount of evidence that the correspondence is correct, a number of very basic questions remain, such as: 
\begin{itemize}
	\item Why and how exactly space-time should emerge from quantum field theories? 
	\item How space-time geometry is encoded in the quantum field theory state?
\end{itemize}
Attempts to answer these questions led to the realization that gravity, as space-time geometry, is intimately related to the entanglement dynamics of a quantum field theory state \cite{Maldacena2,Swingle,Raamsdonk,Ryu-Takayanagi}.

Recently, an heuristic answer to the first question was provided, following a distinct route from that of AdS/CFT correspondence. Central to this progress is the development of a new localization scheme for scalar fields \cite{kiosses}.
The same formulation was already applied in the derivation of Unruh effect in quantum space-time \cite{C-K}.
The method consists of attaching to the momentum space of the free Klein-Gordon field an extra degree of freedom, associated to the vacuum energy, and then defining in the extended momentum space a second field.
Roughly speaking, the new field integrates the, non-interacting, vacuum energy with the total energy of the free Klein-Gordon field, but in practice, it provides, through a well-defined set of time and space operators, the space-time where the standard field is localized. 

In this work, which is a continuation of \cite{kiosses}, making use of the same localization scheme, we will try to give an answer to the second question.
More specifically, we will argue that the geometry of the space-time where the quantum states of the scalar field are localized is determined by the quantum entanglement of these states.
We begin by briefly summarizing the main elements of localization scheme (section \ref{Background}). In \ref{Localization of entangled particle states} we apply the scheme to localize entangled quantum states and in \ref{space-time geometry of entangled states} we show that the derived space-time is curved.

We consider quantum field theories in two dimensions with metric signature $(\text{time, space})=(-,+)$. Furthermore, the units are chosen such that $c = \hbar=1$.

\section{Background}\label{Background}
We consider the simplest possible quantum field theory, a real, scalar field $\Phi$ obeying the Klein-Gordon equation
\be
\left(\partial^2_\ttt - \partial^2_x + m^2\right)\Phi(\ttt,x) = 0. \label{Klein-Gordon_equation}  
\ee
The field can be (second-)quantized by forming the field operator $\hat{\Phi}(\ttt,x)$ in the usual way:
\be
\hat{\Phi}(\ttt,x) = \int d\tp \left(\hat{a}_{\tp} \phi_{\tp}(\ttt,x) + \hat{a}^\dagger_{\tp} \phi_{\tp}^*(\ttt,x) \right), \label{standard_K-G}
\ee
where $\phi_{\tp}(\ttt,x) = e^{-i(E_{\tp} \ttt - \tp x)}/\sqrt{4 \pi E_{\tp}}$
is the normal mode solutions of (\ref{Klein-Gordon_equation}) and $\hat{a}_{\tp}$ ($\hat{a}^{\dagger}_{\tp}$) is the annihilation (creation) operator of the theory, with the associated vacuum state $\ket{0}$ defined by $\hat{a}_{\tp}\ket{0}=0$.
The energy (i.e. Hamiltonian) operator of the theory reads
\be
\hat{H}=\int d\tp \, E_\tp \,\left(\hat{a}_\tp^\dagger\, \hat{a}_\tp +\frac{1}{2} [\hat{a}_\tp, \hat{a}_\tp^\dagger]\right).\label{KG-Hamiltonian}
\ee
For each value of momentum $\tp$, it describes the total energy of the field distributed over various particle states with energy $E_{\tp}= \sqrt{\tp^2 + m^2}$.
The total particle number operator, defined by 
\be
\hat{\tN} = \int d\tp \,\hat{a}_\tp^\dagger\, \hat{a}_\tp,\label{particle_number_operator}
\ee 
commutes with the Hamiltonian $\hat{H}$ and therefore the number of particles is a constant.

In \cite{kiosses} the author has proposed a new localization scheme for Klein-Gordon theory by utilizing its vacuum energy.
More specifically, it was introduced an interaction between the field's standard momentum $\tp$ and vacuum energy $\tE$ in the form of a second field $\wtPsi(\tE,\tp)$, which satisfies the differential equation
\be
\left(\partial_{\tE}^2-\partial_{\tp}^2 - \frac{1}{\kappa^2} \right)\wtPsi(\tE,\tp) = 0, \label{accelerated_diff.-equation}
\ee 
$\kappa$ is a field parameter. Since $\wtPsi$ lives in the (``extended-'') momentum space of Klein-Gordon field $\Phi$, the two fields should be correlated. Indeed, it has been shown that in the solution of (\ref{accelerated_diff.-equation}), the annihilation/creation operators of the new field are represented by the Klein-Gordon field operator and its conjugate transpose:
\be
\hat{\wtPsi}(\tE,\tp) = \int d\ttt \left( \hat{\Phi}_{x}(\ttt) \, \tu_{\ttt}(\tE,\tp) + \hat{\Phi}_{x}^\dagger(\ttt) \, \tu_{\ttt}^*(\tE,\tp)\right),
\label{Field_expansion-Fourier-K_G-Acc}
\ee
where $\tu_{\ttt}(\tE,\tp) = e^{i(\ttt \,\tE - \tx_t\, \tp)}/\sqrt{4 \pi \tx_t}$ the normal mode solutions of (\ref{accelerated_diff.-equation}) with $\tx_t = \sqrt{\ttt^2 + 1/\kappa^2}$.

Following the procedure of canonical quantization, the formal structure of standard field theory is preserved by considering the ``Hamiltonian''
\be
\hat{X} = \int d\ttt \,  \tx_t\, \hat{\Phi}_{x}^\dagger(\ttt)\,\, \hat{\Phi}_{x}(\ttt),
\ee
and the operator
\be
\hat{T} = \int d\ttt\,  \ttt\,\,\hat{\Phi}_{x}^\dagger(\ttt)\,\, \hat{\Phi}_{x}(\ttt) . 
\ee 
Recovering the speed of light constant $c$, it is found that eigenvalues of $\hat{X}$ has dimensions of space (length), while eigenvalues of $\hat{T}$ dimensions of time. However, for reasons of simplicity we continue with units allowing $c=1$.

The utility of space-time operators $\hat{T}^\mu=(\hat{T},\hat{X})$ becomes apparent once we define the $\tN-$particle state vector 
\be
\ket{\tN}:= \left(\hat{\Phi}^\dagger(\ttt,x)\right)^{\tN}\ket{0}. \label{multi-particle states}
\ee
This state is a superposition of $\tN$-particle momentum eigenstates (one realizes that just applying expansion (\ref{standard_K-G})) and thus describes a system of $\tN$ Klein-Gordon particles.
Particle states of this form are well defined entities and conserved in the sense that are eigenstates of both $\hat{H}$ and $\hat{\tN}$ operators. 

Considering the action of $\hat{T}^\mu$ on $\ket{\tN}$, it is found
\be
\hat{T}^\mu\ket{\tN} = t^\mu\tN \ket{\tN}, \label{space-time_eigenequation}
\ee
where $\ttt^\mu:=(\ttt,\tx_t)$.
So, the particle state $\ket{\tN}$ is an eigenstate of space-time operators $\hat{T}^\mu$ with eigenvalues $t^\mu\tN$, a result that implies that $\tN$ Klein-Gordon particles are localized on space-time with coordinates $t^\mu$.
This space-time assumes relative space and time since it accommodates the Lorentz transformations
\be
\tx'_t = \gamma (\tx_t - \tw\ttt)\qquad \ttt' = \gamma (\ttt - \tw\tx_t)
\ee
between two reference frames with coordinates $\ttt'^\mu$ and $\ttt^\mu$ and relative velocity $\tw=d\tx_t/d\ttt$ (for more details see \cite{kiosses}). $\gamma=\sqrt{1-\tw^2}$ is the Lorentz factor. 
It is of crucial importance to notice the scale invariance of field theory (\ref{accelerated_diff.-equation}) which provides us with the space-time $\ttt-\tx_t$. Under a dilation $t^\mu\rightarrow\tN\, t^\mu$, space-time operator $\hat{T}^\mu$, by definition, acquire a factor, the Klein-Gordon number operator $\hat{\tN}=\int d\ttt \hat{\Phi}_{x}^\dagger(\ttt)\,\, \hat{\Phi}_{x}(\ttt)$, which is equivalent to (\ref{particle_number_operator}), see \cite{kiosses}.
Typically, scale invariance indicates that no fixed length scales appears. In subsection \ref{space-time geometry of entangled states}, we shall show how this leads to the definition of curved space-time.

Due to the nature of quantum mechanics we can also get linear combinations of particles states with different number of particles:
\be
\ket{\tL}=\sum_{i} \, \tC_i \ket{\tN_i},
\ee 
where $\sum_i |\tC_i|^2=1$.
$\ket{\tN_i}$ describes a free field theory with $\tN_i$ particles.
In that case it holds
\ba
\langle \hat{\tN}\rangle_\tL&:=&\bra\tL \hat{\tN} \ket{\tL} = \sum_i p_i \tN_i \\ 
\langle \hat{T}^\mu\rangle_\tL &:=&\bra\tL \hat{T}^\mu \ket{\tL}= \sum_i p_i \, (t^\mu \tN_i)
\ea
where $p_i=|\tC_i|^2$.
The probabilities appeared in the expectation values represent some classical uncertainty about the particle state and consequently about its space-time location.
This uncertainty disappears for distinct states like $\ket{\tN}$.

\section{Results}
\subsection{Localization of entangled particle states}\label{Localization of entangled particle states}
Let us consider now a more complicated quantum system by taking two copies of these multi-particle systems and attempt to apply our localization scheme on this.
The Hilbert space of the entire system is decomposed as the tensor product of the Hilbert spaces for the component systems $\mathcal{H}=\mathcal{H}_A \otimes \mathcal{H}_B$.
Given the basis of orthogonal states $\{\ket{\tN_i}\}$ for subsystems $A$ and $B$, we can represent the state of the entire system as
\be
\ket{\tM}:=\sum_{i,j} \tC_{i,j}  \ket{\tN_i}\otimes \ket{\tN_j}
\ee
where the complex coefficients $\tC_{i,j}$ satisfy the normalization condition $\sum_{i,j}|\tC_{i,j}|^2=1$.

State $\ket{\tM}$ can be entangled.
A simple quantitative and convenient way to measure the entanglement between $\mathcal{H}_A$ and $\mathcal{H}_B$ is provided by entanglement entropy (or von Neumann entropy) \cite{N-C}.
The reduced density matrix $\rho_A$ ($\rho_B$) for the subsystem $\mathcal{H}_A$ ($\mathcal{H}_B$) is obtained by taking the partial trace over the subsystem $\mathcal{H}_B$ ($\mathcal{H}_A$) \cite{Levay}
\be
\rho_A := \mathrm{Tr}_B \rho = \tC \tC^{\dagger},\qquad \rho_B := \mathrm{Tr}_A \rho =  (\tC^{\dagger} \tC)^T.
\ee 
Here, $\rho = \ket{\tM}\bra{\tM}$ is the total density matrix and $\tC$ is a matrix with entries $\tC_{i,j}$. 
For the $l$-subsystem, entanglement entropy  is defined as
\be
\begin{aligned}
\tS_l &= -\mathrm{Tr} \rho_l \log \rho_l \\
&= - \left(\lambda_+^{(l)} \log \lambda_+^{(l)} + \lambda_-^{(l)} \log \lambda_-^{(l)}\right)
\end{aligned}
\qquad l=A,B
\ee 
where $\lambda_\pm^{(l)} = \left(1\pm\sqrt{1-(\mathrm{det}\tC)^2}\right)/2$ the eigenvalues of the reduced density matrices.
The $l$-subsystem is maximally entangled with the rest of the system when $\mathrm{det}\tC =1$, while the entanglement entropy vanishes and the subsystem reduces to a separable state, when $\mathrm{det}\tC=0$.

For the sake of clarity of our argument, let us confine ourselves to the case of 
Hilbert spaces of dimension $d=2$ for each subsystem. Then, the entire system can be described by the quantum state
\be
\ket{\tM(\tb)}:=\sum_{\tN_1,\tN_2=\{\tN,\tN-\tb\}} \tC_{\tN_1,\tN_2} \ket{\tN_1}\otimes \ket{\tN_2},
\quad 
-\infty < \tb \le \tN
\ee
Written the general state in that form, parameter $\tb$ varies the quantum state in such a way that the entanglement entropy assigned to it changes. 
Explicitly this becomes apparent once we calculate $ \mathrm{det}[\tC_{\tN_1,\tN_2}]$ ($[\tC_{\tN_1,\tN_2}]$ a matrix with entries $\tC_{\tN_1,\tN_2}$) for various values of $\tb$. We recognize two distinct cases\footnote{of course we can also quantify the degree of entanglement for a subsystem, but this is beyond the purposes of this work.}:
\be
\begin{aligned}
 \mathrm{det}[\tC_{\tN_1,\tN_2}]&=0 \quad \text{for}\quad \tb=0 \\ \mathrm{det}[\tC_{\tN_1,\tN_2}]&\neq0 \quad \text{for}\quad \tb\neq0 \\
\end{aligned}
\ee
This result clearly states that, according to the definition of entanglement entropy we gave above, while the entanglement of state $\ket{\tM(\tb)}$ is not zero, entanglement of $\ket{\tM(0)}$ is zero allowing us to write the latter as the product state:
\be
\ket{\tM(0)} := \ket{\tN}\otimes \ket{\tN}.
\ee

Calculating the expectation value of number operator $\hat{\tN}$ for the state $\tM(\tb)$ we get:
\be
\langle \hat{\tN}(\tb) \rangle := \bra{\tM(\tb)} \hat{\tN} \ket{\tM(\tb)} = \sum_{i=1}^3 p_i \bar{\tN}_i(\tb)
\ee
where
\be
\begin{aligned}
\bar{\tN}_1(\tb)&:=2 \tN,\\
\quad \bar{\tN}_2(\tb)&:=2\tN-\tb,\\
\quad \bar{\tN}_3(\tb)&:=2\tN-2\tb.
\end{aligned}
\ee
A field theory describing $\bar{\tN}_1$ particles, i.e. both subsystems are in quantum state $\ket{\tN}$, occurs with probability $p_1=|\tC_{\tN,\tN}|^2$. A field theory which describes $\bar{\tN}_2$ particles, i.e. one subsystem in state $\ket{\tN}$ and the other in state $\ket{\tN-\tb}$, has probability $p_2=\left(|\tC_{\tN,(\tN-b)}|^2+|C_{(\tN-b),\tN}|^2\right)$. And finally the probability for $\bar{\tN}_3$ particles, which corresponds to quantum states $\ket{\tN-\tb}$ for both subsystems, is $p_3=|\tC_{(\tN-b),(\tN-b)}|^2$.

Passing to expectation values of space-time operators $\hat{T}^\mu$ in the state $\ket{\tM(\tb)}$ we take
\be
\langle \hat{T}^\mu(\tb) \rangle:=\bra{\tM(\tb)}\hat{T}^\mu \ket{\tM(\tb)}=\sum_{i=1}^3 p_i\,  \bar{T}^\mu_i(\tb)\label{space-time-exp1}
\ee
where
\be
\bar{T}^\mu_i(\tb) := \ttt^\mu\,\bar{\tN}_i(\tb).\label{space-time-exp2}
\ee
The field theory for $\bar{\tN}_1$ particles, or the general particle state $\ket{\tN}\otimes\ket{\tN}$, is localized in space-time position $\bar{T}^\mu_1$ with probability $p_1$. $\bar{\tN}_2$ particles, described by the particle states $\ket{\tN}\otimes\ket{\tN-\tb}$ or $\ket{\tN-\tb}\otimes\ket{\tN}$, are located at space-time position $\bar{T}^\mu_2$ with probability $p_2$. And lastly, the position $\bar{T}^\mu_3$ is assigned to the field theory describing $\bar{\tN}_3$ particles, or general particle state $\ket{\tN-\tb}\otimes\ket{\tN-\tb}$, with probability $p_3$.

As expected, the uncertainty on number of particles and space-time localization fades away for unentangled particle states, i.e. in case of $\tb=0$,
\be
\langle \hat{\tN}(0) \rangle =\bar{\tN}_i(0) =2\tN\quad \text{and}\quad \langle \hat{T}^\mu(0) \rangle=\bar{T}^\mu_i(0)=2 t^\mu \tN.\label{eq.25}
\ee

In the next subsection, we will show now that the unentangled state $\ket{\tM(0)}$ is localized, with certainty, on flat space-time, while the entangled state $\ket{\tM(\tb)}$, with some uncertainty, on curved space-time.

\subsection{space-time geometry of entangled states}\label{space-time geometry of entangled states}

Our aim here is to investigate the geometry of space-time made of the totality of all expectation values $\langle \hat{T}^\mu(\tb) \rangle$.
Mathematically an effective way to study the geometry of a space-time is analyzing its metric.
Let us start, however, with the calculation of the interval between two arbitrary space-time points before we derive the line element of the space-time.
Due to eqs.(\ref{space-time-exp1},\ref{space-time-exp2}), we realize that the actual points of our space-time are the eigenvalues $\bar{T}_i^\mu$, keeping of course in mind that each is weighted with some probability and assuming the coordinate system $\ttt^\mu$ for space-time measurements.  
In order to keep our notation as simple as possible we choose to calculate the distance of $\bar{T}_i^\mu$ from the origin $\ttt^\mu=(0,0)$.
Making use of the relations (\ref{space-time-exp2}) and $\tx_t^2-\ttt^2=1/\kappa^2$ we obtain
\be
\bar{S}^2_i(\tb):= - \left(\bar{T}^0_i(\tb)\right)^2 + \left(\bar{T}^1_i(\tb)\right)^2 =\left(\frac{\bar{\tN}_i(\tb)}{\kappa}\right)^2.\label{bsb2}
\ee
The interval $\bar{S}^2_i(\tb)$, an arbitrary interval in space-time $\bar{T}_i^\mu$, is independent of coordinate system in which it is calculated, but does depend on the number operator's eigenvalues $\bar{\tN}_i$.
This result actually verifies our earlier assertion on the scale invariance of field theory (\ref{accelerated_diff.-equation}).

In classical relativistic physics, space-time interval $S^2$ is of great importance because gives an invariant geometrical meaning to certain physical statements.
For instance, a negative defined interval $S^2<0$ between two space-time points is the time elapsed on a clock which starts from the one and passes through the other point. While, a positive defined interval, $S^2>0$, is the length of a ruler that joins simultaneously two space-time points \cite{Schutz}.

From that perspective  the dependence of $\bar{S}^2_i(\tb)$ on the outcome of $\langle \hat{\tN}(\tb)\rangle$ renders coordinates $\bar{T}^\mu_i(\tb)$ (and consequently expectation values $\langle \hat{T}^\mu(\tb) \rangle$) physically meaningless, in the sense that these measurements cannot be delivered by any real clock and any real ruler. 
But this is not the case for the space-time $\left\{\langle \hat{T}^\mu(0)\rangle\right\}$, which we obtain from the set of all localization points of product state $\ket{\tM(0)}$. On this occasion, because of eqs.(\ref{eq.25}), it corresponds a fixed space-time interval, 
\be
S^2(0):= - \langle \hat{T}^0(0)\rangle^2 + \langle \hat{T}^1(0)\rangle^2=\left(\frac{2\tN}{\kappa}\right)^2,\label{so2}
\ee
and thus, contrary to the space-time of entangled state $\ket{\tM(\tb)}$,  an absolute structure that can accommodate length and time measurements.

Space-time $\left\{\langle \hat{T}^\mu(0)\rangle\right\}$ can be incorporated to the more abstract space-time $\left\{\langle \hat{T}^\mu(\tb)\rangle\right\}$ by the prescription of spacetime intervals and the fact that both space-times are products of the same field theory. 
More specifically, both spacetime intervals, $\bar{S}^2_i(\tb)$ and $S^2(0)$, were found that it is inverse proportional to the field parameter $\kappa$.
Combining equations (\ref{bsb2}) and (\ref{so2}) we get:
\be
\bar{S}^2_i(\tb):=\left(\frac{\bar{\tN}_i(\tb)}{2\tN}\right)^2 S^2(0) .
\ee
Thus, with the help of $S^2$, the generalized spacetime intervals $S^2_i$ are physically defined by lengths of rulers and readings of clocks, and from
\be
\bar{T}^\mu_i(\tb) = \left(\frac{\bar{\tN}_i(\tb)}{2\tN}\right)\, T^\mu \label{eq.34}
\ee
every space-time point transforms from $\bar{T}^\mu_i(\tb)$ to the physically meaningful coordinates $T^\mu$ (From now on, to reduce clutter in our notation we define $T^\mu:=\langle \hat{T}^\mu(0)\rangle$).
Of course, as expected, by disentangling the particle state, i.e. $\tb=0$, the two coordinate systems coincide, $\bar{T}^\mu_i(0)=T^\mu$.

Having physically identified the generalized coordinates $\bar{T}^\mu_i(\tb)$, we can proceed now to the derivation of the metric $\bar{g}_{\mu\nu}^i(T^\mu)$, as a set of functions of positions, such that the line element 
\be
d\bar{S}^2_i=\bar{g}_{\mu\nu}^i(T^\lambda) \, dT^\mu dT^\nu,\qquad \lambda,\mu,\nu=0,1 \label{line_element}
\ee
expresses the interval between $\bar{T}^\mu_i(\tb)$ and $\bar{T}^\mu_i(\tb)+d\bar{T}^\mu_i(\tb)$.

An interesting property of the space-time we study is that it is stationary. The Klein-Gordon field theory we consider here is free, so $[\hat{\tN},\hat{H}]=0$ and by definition $[\hat{\tN},\hat{T}^\mu]=0$. These ensure that the spectrums of number operator $\hat{\tN}$ and space-time positon operator $\hat{T}^\mu$ are constant in time.
With this in mind we assume that there exists a rigid space-filling \textit{lattice} constructed out of massless rigid rulers, and that allows us to identify space points fixed in the space-time with lattice points. 

Another compelling property our derived space-time acquires is that it admits a preferred time.
Classically we know that the preferred time is the result of a sensible synchronization prescription \cite{Rindler}. This prescription can be pictorially described as follows. We imagine clocks, attached to all the lattice points. These clocks are furnished with a lever allowing their standard rates to be re-adjusted by an arbitrary constant factor.
One of these clocks is chosen to be the master clock, and is allowed to tick at its standard rate. All the other clocks are then rate-synchronized with the master clock.

We apply the same synchronization prescription to our case. 
We assume that the clocks measure time $T^0$ and are attached to the lattice points with coordinates $\bar{T}^1_i(\tb)$. 
Then for successive readings of clock at position $\bar{T}^1_i(\tb)$ we get
\be
d\bar{S}_i^2(\tb) = -\left(\frac{\bar{\tN}_i(\tb)}{2\tN}\right)^{-2} \,(dT^0)^2,\qquad \bar{T}^1_i(\tb)=\text{const.} \label{static_st-cons}
\ee 
where, due to eq.(\ref{eq.34}), in front of $dT^0$ we have introduced the factor by which the clock rate at $\bar{T}^1_i(\tb)$ changes.

The metric will be, as always, a quadratic form in the coordinate differentials. But choosing the space-time to be static, our synchronization procedure implies the vanishing of the time–space cross-terms \cite{Rindler}. So the general line element takes the form 
\be
d\bar{S}^2_i(\tb) = -\left(\frac{\bar{\tN}_i(\tb)}{2\tN}\right)^{-2} \,(dT^0)^2 +(d\bar{T}^1_i(\tb))^2. \label{static_st1}
\ee 
The detailed form of $dT^1_i$ is not unknown to us, it is given by (\ref{eq.34}). So finally we get
\be
d\bar{S}^2_i(\tb) = -\left(\frac{\bar{\tN}_i(\tb)}{2\tN}\right)^{-2}(dT^0)^2 +\left(\frac{\bar{\tN}_i(\tb)}{2\tN}\right)^2(dT^1)^2 \label{static_st}
\ee
or, expressed in the compact form of eq.(\ref{line_element}) with
\be
\bar{g}_{\mu\nu}^i =
\left(
\begin{matrix}
 -\left(\frac{\bar{\tN}_i(\tb)}{2\tN}\right)^{-2} & 0 \\
 0 & \left(\frac{\bar{\tN}_i(\tb)}{2\tN}\right)^{2} \\
\end{matrix}
\right).
\ee
That metric $\bar{g}_{\mu\nu}^i$ is a set of functions of position, i.e. $\tN=\tN(T^1)$ and $\tb=\tb(T^1)$. The latter follows from the commutation relation $[\hat{\tN},\hat{T}^1]=0$.
However, for later convinience we express it just as function of the entanglement parameter $\tb$, $\bar{g}_{\mu\nu}^i=\bar{g}_{\mu\nu}^i(\tb)$.

The metric we have obtained depends on the square of the outcome $\bar{\tN}_i$, the eigenvalue of the number operator $\hat{\tN}$.
Substituting $\bar{\tN}_i(\tb)^2$ with its expectation value $\left\langle\bar{\tN}_i(\tb)^2\right\rangle=\sum_i p_i \bar{\tN}_i(\tb)^2$ apparently we get the expectation value of metric
\be
\langle g_{\mu\nu}(\tb)\rangle  = \sum_{i=1}^3 P_i  \,\bar{g}_{\mu\nu}^i(\tb)
\ee
where
\be
P_i:=\left(
\begin{matrix}
	\frac{1}{p_i} & 0 \\
	0 & p_i \\
\end{matrix}
\right).
\ee
The metric we derived for a general entangled quantum state $\ket{\tM(\tb)}$ possess simultaneously, both properties of gravitational physics, i.e. curvature, and quantum mechanics, i.e. probability.
The metric coefficients signify the deviation from flat-spacetime values $\text{diag}(-1,1)$ and thus the emergence of curved space-time.
Additionally, space-time measurements or metric outcome $\bar{g}_{\mu\nu}^i(\tb)$, cannot be predicted with perfect confidence, even in principle. One can only calculate the probability of obtaining each possible outcome.

However, the same cannot be said for the seperable state $\ket{\tM(0)}$. 
For $\tb=0$ the expectation value of metric reduces to
\be
\langle g_{\mu\nu}(0)\rangle  = \left(
\begin{matrix}
1 & 0 \\
	0 & 1 \\
\end{matrix}
\right),
\ee
since in that case holds $\left\langle\bar{\tN}_i(0)^2\right\rangle = \left\langle\bar{\tN}_i(0)\right\rangle^2 =(2\tN)^2$.
The metric, with certainty, is that of a flat space-time.

\section{Conclusions}

We have seen that the localization scheme introduced in \cite{kiosses} establishes a connection between entanglement of ordinary matter states and gravitation as curved space-time.
While the space-time where an entangled particle state is localized can be curved, as the entanglement drops to zero, the space-time definitely becomes flat.

\noindent
\section*{Acknowledgments} 
I acknowledge financial support from Instituto Nazionale di Ottica - Consiglio Nazionale delle Ricerche (CNR-INO).

\end{document}